# The IAEA Fusion Data Lake Project - Accelerating AI and Big Data Applications through Open Science and FAIR Data

DS Gahle[1] and M Barbarino[1*]
1. International Atomic Energy Agency, Vienna, Austria
*Email: m.barbarino@iaea.org

## Introduction: Data Infrastructure Needs for Accelerating Fusion Development with AI

AI applications are accelerating technology progress and deployments across fields, including fusion. The use of neural networks as surrogate models and digital twins and large language models for experiment log analytics and data model developments is facilitating advances in scenario design, plasma control, and physics insights [1-3]. Underpinning these activities is a need for modern data platforms and metadata-rich data sets. Current efforts to scale up big data and AI/ML activities are limited by the availability of such datasets, with many institutions, such as the UK Atomic Energy Agency [4] and Japan's National Institute for Fusion Science [5], increasing efforts to build out the required infrastructure. The IAEA has a unique role to play by facilitating collaboration across the global fusion scientific community, industry, and academia. A key IAEA activity in this space is the AI for Fusion Coordinate Research Project (CRP), a five-year initiative launched in 2022, which brings together 22 institutions across 11 countries. It takes a holistic view covering physics and engineering applications in magnetic and inertial fusion energy, underlying data infrastructure, and workforce development [6]. Within this CRP is the Fusion Data Lake (FDL) project, leveraging the position of the IAEA to further data sharing and integration globally and developing the modern data infrastructure needed to provide the multimachine datasets needed to develop machine-agnostic models that can scale to future fusion devices that follow the FAIR data principles, with both a human-friendly interface and a programmable API for integrating in analysis pipelines.

## Platform Architecture and Technologies

For the project there are three main service goals to deliver: (1) Data Catalogue - A query-able central catalogue of experiments and diagnostic signals from devices around the world; (2) Data Federation - Lower the barries to data access and sharing by producing a single access point to a decentralised network of experimental and simulation repositories of different institutions; and (3) Centralised storage - Provide interim short to medium term data storage for unfederated data.

The platform architecture has been designed with these goals in mind on top of the enterprise technology stack for the IAEA for modernising data solutions: Snowflake for the computation resource of Python data pipeline and SQL database servers; Microsoft Azure for storage and Azure DevOps for the CI/CD platform; and finally C# and .Net for webapp development. The data pipeline structure takes the conventional ETL (Extract, Transform, Load) approach within the modern medallion structure of data storage, as illustrated in Figure 1 below showing the relation between the data sources, pipeline infrastructure, and user interfaces. A default metadata-driven ingestion pipeline has been developed; using a configuration file to manage the transformation of the datasets (including mapping to the data model) and defining the target databases and tables from the configuration. This structure is reusable, reducing the effort expected for maintenance and ingestion of new sources.

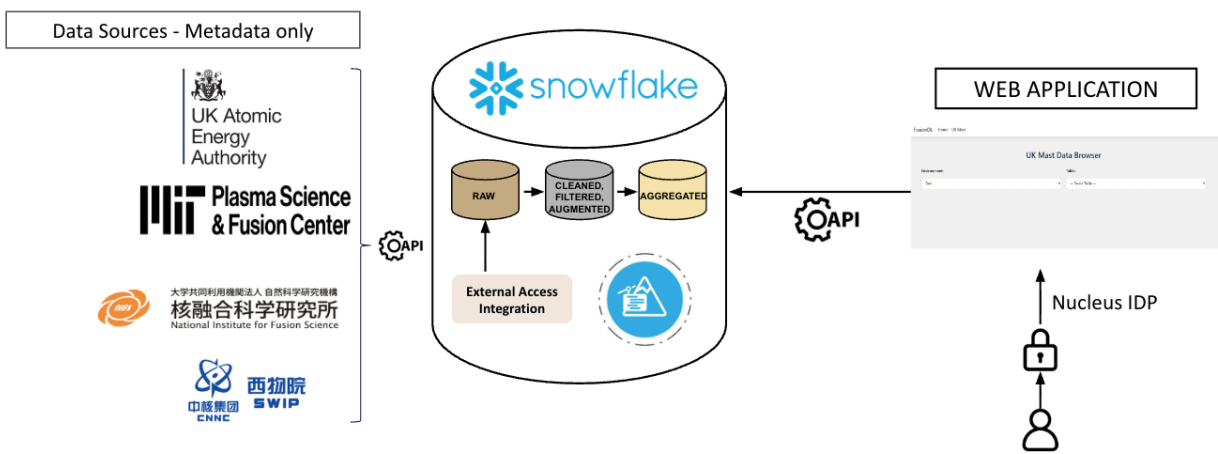

Figure 1. Shows the high level architecture of the Fusion Data Lake: from the various data sources; through to ETL ingestion pipeline to the medallion structured catalogue databases; to the web application user interface with the user access pathway.

An initial data model has been created for the high-level catalogue for the platform: the Minimal Metadata Model [7]. This simple model lists the simplest of metadata of interest from a fusion experiment, categorised over three tables: "core", "icf", and "mcf". The ontology for this model is being developed in line with the ITER IMAS Data Dictionary to ease the adoption of the Fusion Data Lake [8].

## Proof of Concept: Phases I, II, & III

The development of the project is evolving over three proof of concept (PoC) phases to build the initial product for early access (live) testing: Phase I - demonstrate the data federation concept; Phase II - demonstrate the scalability of the implementation; and Phase III - expand the diverse datasets available and prepare for the pre-release of the platform. Currently, Phase II is being completed, and preparatory work is being undertaken to commence Phase III.

Phase I built the initial data platform cover the data ingestion pipeline and storage within the Snowflake cloud platform, with a simple web application that displayed the data catalogue in an interactive table. The data ingested was the Mega Ampere Sphereical Tokamak (MAST) shot log from UKAEA's MAST Data Catalog [4, 9], using the REST API developed by UKAEA. This demonstrated the federation concept, though synchronising the Fusion Data Lake and MAST Data Catalog catalogues, and enabling FDL users to download MAST data from the platform without that data going through the FDL infrastructure.

Phase II is demonstrating the scalability of the platform by adding the data catalogues of the Large Helical Device (LHD) of the National Institute for Fusion Science in Japan [10], and Alcator C-Mod of the Plasma Science and Fusion Center at Massachusetts Institute of Technology (PSFC MIT) in the USA [11]. The platform's scalability is demonstrated by the integration of three different datasets (MAST, LHD, and Alcator C-Mod) conforming to a single FDL data model and ingested through unique patterns. For the LHD dataset, data was shared through the IAEA Microsoft Azure Blob Storage, which is then synchronised with the Snowflake platform; and ingestion of the Alcator C-Mod dataset is being ingested through a combination of PSFC tools (disruption-py) [12], MDS Plus [13], and the aforementioned IAEA's multicloud platform.

Phase III will further expand by incorporating the catalogue of HL-2A of the Southwestern Institute of Physics in China [14]. Beyond catalogue expansion, the front-end interfaces (both API and web applications) will be developed to a level of maturity that will enable early access users to test the platform in a manner compliant with the platform Terms of Service.

## Provisional Data Governance Strategy

Key to the success of international enterprise-level data projects is the data and platform governance that regulates the platform, the data providers, and the data users, both for the purposes of developing a service that is scalable in operations and business-to-business integration. To that end, provisional regulations are in development to facilitate this project and prepare for a soft launch of a pre-release version of the platform. Initially, this is comprised of the Terms of Service and data access-privacy levels.

Terms of Service (ToS) are foreseen to cover at least three main areas: uploads-ingress, downloads-egress, and attribution-citation. For data uploads, users must have (and declare) that they have the authority to share the respective data with the platform; and, the data must be licenced, which is enriched in the metadata directly or linked. Downloaded data must only be used and shared in line with the Fusion Data Lake platform ToS and data licensing associated with each individual data object. The platform will facilitate transparency of the data licensing and, in that way, enable the data user compliance. Data must be appropriately cited and acknowledged in line with the respective data licenses, and citation of the Fusion Data Lake platform when used as a data service in a project or work.

Data access and privacy level are being developed to enable institutions and member states to share data in line with their data licensing and organisational policies. In order of user accessability, the access-privacy levels are: (1) Public - Data is publicly accessible without login credentials; (2) Internal - Data is accessible to any user with an IAEA NUCLEUS account (self-service account creation)[15]; (3) Restricted - Data only available to users with NUCLEUS credentials from institutions of types of institutions preapproved by the data owner (possible credential authentication using OpenAthens [16]); and (4) Closed - Data only available to users with NUCLEUS credentials who have been individually approved by the data owner. As the project develops, a matrix of permissions will be developed for data to cover the rights to view, download, and edit the data.

## Conclusions and the Future

The Fusion Data Lake project is leveraging the unique position of the IAEA to develop a globally inclusive modern data platform that supports the needs of the fusion community for metadata-rich multimachine datasets on infrastructure that is both human accessible and programmable into modern machine learning pipelines, a foundation of modern AI/ML work. The evolution of the Proof of Concept over three phases has demonstrated the scalability and suitability of the platform to the task, as well as the technological and governance ability to integrate a diverse range of data sets and institutions. Going forward, the platform will continue to work with partners within the AI for Fusion CRP and without to expand the offering and capability of the Fusion Data Lake and prepare for pre-release.

## Acknowledgements

The authors would like to thank Fudan University and the Programme Committee of the Sixth IAEA Technical Meeting on Fusion Data Processing, Validation, and Analysis for the opportunity to share this work with the community. This report is based on the work presented at the meeting. The authors would also like to thank the institutes that are part of the IAEA AI for Fusion CRP, who have enabled and supported this work. This work was developed and completed with IAEA funding. The presented work reflects only the views of the authors and may not represent any affiliated organisations.

## Institutes of the IAEA AI for Fusion CRP

The full list of institutions and partners of the IAEA AI for Fusion CRP is: Rafael Bordas (First Light Fusion Limited, United Kingdom), Aidan Crilly (Imperial College London, United Kingdom), Eliana De Marchi (Eni S.p.A., Italy), Charles Gretton (Australian National University, Australia), Jeongwon Lee (Korea Institute of Fusion Energy, Republic of Korea), Ryan McClarren (University of Notre Dame, United States of America), Saskia Mordijck (College of William and Mary, United States of America), Hideo Nagatomo (Institute of Laser Engineering, Osaka University, Japan), Alessandro Pau (Swiss Plasma Center, Switzerland), Stanislas Pamela (United Kingdom Atomic Energy Authority, United Kingdom), Giuseppe A. Ratta (Laboratorio Nacional de Fusión por Confinamiento Magnético, Spain), Cristina Rea (Massachusetts Institute of Technology, United States of America), Brian Sammuli (General Atomics, United States of America), Jaemin Seo (Chung-Ang University, Republic of Korea), Manika Sharma (Institute for Plasma Research, India), Giuliana Sias (Università degli Studi di Cagliari, Italy), David Smith (University of Wisconsin–Madison, United States of America), Par Strand (Chalmers University of Technology, Sweden), Fuyuan Wu (Shanghai Jiao Tong University, China), Bingjia Xiao (Chinese Academy of Sciences – Institute of Plasma Physics, China), Masayuki Yokoyama (National Institute for Fusion Science, Japan), Wulyu Zhong (Southwestern Institute of Physics, China).